\documentclass{article}

\usepackage{arxiv}

\usepackage[utf8]{inputenc} % allow utf-8 input
\usepackage[T1]{fontenc}    % use 8-bit T1 fonts
\usepackage{hyperref}       % hyperlinks
\usepackage{url}            % simple URL typesetting
\usepackage{booktabs}       % professional-quality tables
\usepackage{amsfonts}       % blackboard math symbols
\usepackage{nicefrac}       % compact symbols for 1/2, etc.
\usepackage{microtype}      % microtypography
\usepackage{lipsum}

\usepackage[utf8]{inputenc}
\usepackage[english]{babel}
\usepackage[affil-it]{authblk}
\usepackage[utf8]{inputenc}
\usepackage{amsmath}
\usepackage{algorithm}
\usepackage[numbers]{natbib}
\usepackage{adjustbox}
\usepackage{blindtext}
\usepackage[english]{babel}
\usepackage[utf8]{inputenc}
\usepackage{rotating}
\usepackage{natbib}
\usepackage{blindtext, rotating}
\usepackage{tikz}

\usepackage{listings} 
\usepackage{xcolor}
\usepackage{amsfonts}
\usepackage{graphicx}
\usepackage[colorinlistoftodos]{todonotes}
\usepackage{algpseudocode}
\usepackage{geometry}
\usepackage{bbm}
\usepackage{caption}

\title{Formal Modelling  and Verification of Software Defined Network}

\author{
  Jnanamurthy H K and Vijay Varadharajan
 \\ Department of Computer Science\\
  University of Newcastle\\
  Australia \\
  \texttt{jnanamurthy.hk@gmail.com} \\
}

\begin{document}
\maketitle

\begin{abstract}
In cloud computing, software-defined network (SDN) gaining more attention due to its advantages in network configuration to improve network performance and network monitoring. SDN addresses an issue of static architecture in traditional networks by allowing centralised control of a network system. SDN contains centralised network intelligence module which separates a process of forwarding packets (data plane) from packet routing process (control plane). It is essential to ensure the correctness of SDN due to secure data transmitting in it. In this paper. Model-checking is chosen to verify an SDN network. The Computation Tree Logic (CTL) and Linear Temporal Logic (LTL) used as a specification to express properties of an SDN. Then complete SDN structure is defined formally along with its Kripke structure. Finally, temporal properties are analysed against the SDN Kripke model to assure the properties of SDN is correct. 
\end{abstract}

% keywords can be removed

\section{Introduction}
Software-Defined Network (SDN) \cite{SDN} is gaining more attention due to its advantages in network configuration to improve network performance and network monitoring. So it is essential to ensure the correctness of SDN \cite{jnani2}. In this article,  we describe SDN in the Kripke structure along with specifications, then the security properties of an SDN are expressed using temporal formulas. Later, the security properties of the SDN are analysed against a designed Kripke model using model checking. 

There are several formal verification techniques \cite{formal}\cite{MC} studied in recent years such as model checking, abstract interpretation and boolean satisfiability \cite{jnani}\cite{jnani1}.  Model-checking is widely used in many fields such as verification of hardware, software and security and safety protocols \cite{manu}\cite{manu1}. In model checking, the system is modelled as a state machine and specifications (properties of the system) expresses in linear temporal logic and computation tree logic. 

SPIN \cite{SPIN} and SMV \cite{SMV} are famous model checkers, LTL specifications \cite{LTL} are allowed to express in SPIN and CTL specifications \cite{LTL} are in SMV. It is known that automata-based verification can lead to state explosion problem. To address the state explosion problem, symbolic model checking (SMC) \cite{SMC} and abstract model checking (AMC) \cite{AMC} has been developed with successful results. The combination of SMC and Binary Decision Trees (BDDs) \cite{BDD} maximise the states in the system, but a bottleneck in manipulating the amount of memory required to store BDDs. Bounded Model Checking (BMC) \cite{BMC} progresses fastly after SMC, the basic idea of BMC is to find a counterexample in executions whose length k. The BMC problem can quickly reduce to satisfiability problem, which can be verified using SAT solvers \cite{SAT}. Modern SAT solvers can handle satisfiability problem with thousands of variables.

The contribution of the paper is as follows:
\begin{itemize}
    \item  Formal representation of software-defined networks. 
    \item Representation of SDN in a Kripke structure for verification.
    \item Formal analysis of SDN properties against the formal SDN Kripke model.
\item Formal analysis of faulty transition in the Kripke model of SDN.
\end{itemize}

\par The rest of this paper is organised as follows. The necessary background required for the proposed method is defined in Section 2. Section 3 and 4 describes the software-defined network and our solution, including a designed formal model for SDN and verification properties of SDN. Finally, we conclude the work in Section 5.

\section{Background}
\label{background}

\subsection{Kripke Structure}
A Kripke structure M \cite{kripke} is a tuple M=(Init, States, Transition, Labelling), where  `States' is the set of states which are defined by a set of propositions `A' hold on the states, `Init' $\subseteq$ `States' is the set of initial states, `Transition' $\subseteq$ `States' $\times$ `States' and  `Labelling' is a labelling function, `Labelling': `States'$\rightarrow$ 2$^A$.

\subsection{Linear-Time Temporal Logic}
Linear-time temporal logic, or LTL for short, is a temporal logic, with connectives that allow us to refer to the future. It models time as a sequence of states, extending infinitely into the future. This sequence of states is sometimes called a computation path, or simply a path. Linear-time temporal logic (LTL) has the following syntax given in Backus Naur form:
	\begin{center}
	$\phi$::= $\top$ $|$ $\bot$ $|$ p $|$ ($\neg$$\phi$) $|$ ($\phi$ $\wedge$  $\phi$) $|$  ($\phi$ $\vee$  $\phi$) $|$ ($\phi$ $\rightarrow$  $\phi$) $|$ (X$\phi$) $|$ (F$\phi$)  $|$ (G$\phi$) $|$ ($\phi$ U $\phi$)  $|$ ($\phi$ R $\phi$)
	\end{center}

where p is any propositional atom from some set of Atoms. Thus, the symbols $\top$ and $\bot$ are LTL formulas, as are all atoms from Atoms; and $\neg$$\phi$ is an LTL formula if $\phi$ is one, etc. The connectives X, F, G, U and R are called temporal connectives. X means ‘neXt state,’ F means ‘some Future state,’ and G means ‘all future states (Globally).’ The next two, U and R  are called ‘Until’ and ‘Release’  respectively.

\textbf{Definition 1: }In a given model M, a path $\Sigma$ is defined as a sequence of connected edges which connect nodes, lets say $s_{0}$, $s_{1}$, $s_{2}$. . .$s_{n}$ in S are the nodes such that $\forall$m$\geq$0$, s_{m}$ $\longrightarrow$ $s_{m+1}$. The path $\Sigma$= $s_{0}$, $s_{1}$, $s_{2}$. . .$s_{n}$ represents a sequence of nodes in a system M, we define $\Sigma^{i}$ as starting from $s_{i}$, for example $\Sigma^{5}$ is $s_{5}$, $s_{6}$, $s_{7}$. . . $s_{n}$.

\textbf{Definition 2: }A satisfaction relation $\models$ is defined for the model M, considering paths $\Sigma$ to check linear temporal logic (LTL) formula satisfies $\Sigma$. The satisfaction relation $\models$ over $\Sigma$ and LTL formula is specified as:

\begin{itemize}
\item $\Sigma$ $\models$ $\top$, where $\top$ represents true 
\item $\Sigma$ $\models$ q, iff q $\in$ L(s)
\item $\Sigma$ $\models$ X$\psi$ iff $\Sigma^2$ $\models$ $\psi$
\item $\Sigma$ $\models$ G$\psi$ iff $\forall$m  m$\geq$1, $\Sigma^m$ $\models$ $\psi$
\item $\Sigma$ $\models$ F$\psi$ iff $\exists$m m$\geq$1, $\Sigma^m$ $\models$ $\psi$
\item $\Sigma$ $\models$ $\psi_1$ U $\psi_2$ iff $\exists$m  m$\geq$1 such that $\Sigma^m$ $\models$ $\psi_2$ and $\forall$n, n=1, 2, 3...m-1 satisfies $\Sigma^n$ $\models$ $\psi_1$
\end{itemize}

\subsection{Computation Tree Logic}
\textbf{Definition 3:} Computation Tree Logic is a technique to represent time in a tree-like structure in which future is not determined; there  are many paths in which `actual' path is realised. The Backus Naur form of CTL formulas are defined as:
   
    $\phi$ ::= $\top$ $|$ $\bot$ $|$ p $|$ ($\neg$$\phi$) $|$ ($\phi$ $\wedge$  $\phi$) $|$  ($\phi$ $\vee$  $\phi$) $|$ ($\phi$ $\rightarrow$  $\phi$) $|$ (AX$\phi$) $|$ (EX$\phi$) $|$ (AG$\phi$) $|$  (EG$\phi$) $|$ (AF$\phi$)  $|$   (EF$\phi$)  $|$ (A[$\phi$ U $\phi$])  $|$ (E[$\phi$ U $\phi$]) 
   
\textbf{Definition 4:} Let M = (S, $\rightarrow$, L) be a model for CTL, s in S, $\phi$ a CTL
formula. The relation M, s  $\models$  $\phi$ is defined by structural induction on $\phi$:

\begin{itemize}
\item M, s $\models$ $\top$ and  M, s $\not\models$ $\bot$
\item M, s $\models$ q iff q $\in$ L(s)
\item M, s $\models$ $\neg\phi$ iff M, s $\not\models$ $\phi$
\item M, s $\models$ $\phi_{1}\wedge\phi_{2}$ iff M, s $\models$ $\phi_{1}$ and M, s $\models$ $\phi_{2}$
\item M, s $\models$ $\phi_{1}\vee\phi_{2}$ iff M, s $\models$ $\phi_{1}$ or M, s $\models$ $\phi_{2}$
\item M, s $\models$ $\phi_{1}\rightarrow\phi_{2}$ iff M, s $\not\models$ $\phi_{1}$ or M, s $\models$ $\phi_{2}$
\item M, s $\models$ AX$\phi$ iff for all $s_{1}$ such that s $\rightarrow$ $s_{1}$ we have M, $s_{1}$ $\models$  $\phi$. Thus, AX says:
‘in every next state’.
\item M, s $\models$ EX$\phi$ iff for some $s_{1}$ such that s $\rightarrow$ $s_{1}$ we have M, $s_{1}$ $\models$  $\phi$. Thus, EX says:
‘in some next state’.
\item M, s $\models$ AG$\phi$ iff for all paths $s_{1}$ $\rightarrow$ $s_{2}$ $\rightarrow$ $s_{3}$ $\rightarrow$ ..., where $s_{1}$ equals s, and all $s_{i}$ along the path, we have M, $s_{i}$ $\models$ $\phi$.
\item M, s $\models$ AF$\phi$ iff for some paths $s_{1}$ $\rightarrow$ $s_{2}$ $\rightarrow$ $s_{3}$ $\rightarrow$ ..., where $s_{1}$ equals s, and all $s_{i}$ along the path, we have M, $s_{i}$ $\models$ $\phi$.
\item M, s $\models$ AF$\phi$ iff for all paths $s_{1}$ $\rightarrow$ $s_{2}$ $\rightarrow$ $s_{3}$ $\rightarrow$ ..., where $s_{1}$ equals s, and  there is
some $s_{i}$ such that M, $s_{i}$ $\models$ $\phi$.
\item M, s $\models$ EF$\phi$ iff for some paths $s_{1}$ $\rightarrow$ $s_{2}$ $\rightarrow$ $s_{3}$ $\rightarrow$ ..., where $s_{1}$ equals s, and  for some $s_{i}$ along the path, we have M, $s_{i}$ $\models$ $\phi$.
\item M, s $\models$ A[$\phi_{1}$U$\phi_{2}$] iff for all paths $s_{1}$ $\rightarrow$ $s_{2}$ $\rightarrow$ $s_{3}$ $\rightarrow$ ..., where $s_{1}$ equals s, that path satisfies $\phi_{1}$ U $\phi_{2}$.
\item M, s $\models$ E[$\phi_{1}$U$\phi_{2}$] iff for some paths $s_{1}$ $\rightarrow$ $s_{2}$ $\rightarrow$ $s_{3}$ $\rightarrow$ ..., where $s_{1}$ equals s, that path satisfies $\phi_{1}$ U $\phi_{2}$.
\end{itemize}

\section{Software Defined Network}
SDN works on OpenFlow protocol; OpenFlow is a communication interface defined between the controlling plane and data plane.  OpenFlow allows direct access to data plane devices, such as switches and routers. OpenFlow separates network control from networking switches and allows centralized control of a network. OpenFlow allows controller software to define how a network flow passes through the network devices based on application and cloud resources. OpenFlow enables the network to be programmed based on a per-flow basis. An OpenFlow architecture provides granular control on a network to respond to dynamic changes at an application level; this architecture overcomes disadvantages of IP-based routing, which follows the same path regardless of different requirements. Here we present the OpenFlow specification terms:
\begin{itemize}
    \item Action: an operation on a packet to perform forward and modify a packet. 
    \item Connection: a network connection between a switch and a controller. 
    \item Flow table: it contains a set of flow entries. 
    \item  Flow entry: a unit in a flow table to process packets. Flow entry contains match fields for matching packet headers and instructions to apply on a packet. 
    \item Packet forwarding: forwarding packet to an output port or a set of output ports. 
    \item Packet forwarding: forwarding packet to an output port or a set of output ports. 
    \item Header: a control information present in a packet used by a switch to recognise the packet and to inform the switch for forwarding the packet.
    \item Message: a message sent from the OpenFlow protocol for OpenFlow connection. 
\end{itemize}

In this section, we present a formal definition of SDN and security properties which are necessary for SDN. SDN is a tuple SDN:= (P$^{t}$, W, C, FTab, T, ConfiG) where,\\
\subsection{Packet}
   A packet P$^{t}$ is a tuple (h, pLD) consists of a packet header `h' and payload data `pLD' information which is transmitted in a network.  The packet header $<$st, dt, ($\alpha_{1}$, $\alpha_{1}$, . . $\alpha_{n}$)$>$ contains source address `st', destination address `dt' and packet pattern ($\alpha_{1}$, . . $\alpha_{n}$) which is used to match with ports of the switch during transmission in a network system.

\subsection{OpenFlow Switch}
An OpenFlow switche `w' contains one or more flow tables, which performs packet lookup and forwarding. A switch communicates with the controller, and the controller manages the switch via OpenFlow protocol.  The controller can add, delete and update the flow entries of the flow table belongs to switches. Each entry of flow table contains match fields and instructions to match packets. Matching of the packet header with the first flow table continues to additional flow tables. If a matching entry found, then those instructions are used to forward a packet. If no match of entries in the flow table, then a further configuration of a missed flow entry has to proceed by forwarding a packet to the controller, or packet can be dropped. The instructions belong to each entry of flow table contains actions. The actions are the instructions to forward or modify a packet.

`W' is a set of switches W=\{$w_{1}, w_{2}, w_{3}. . . w_{n}$\}, each switch is a tuple w:= (P, W$^c$, FR) where `P' is a set of ports in a switch represented as P=\{$p_{1},p_{2},p_{3}. . .p_{n}$\}. A port `p' either a input or a output port represented as p(ip, op), where `ip' is a set of input ports and `op' is a set of output ports. Each input and output port consists of port ID, which is used to forward a packet and drop a packet based on forwarding rules which are received from the controller. W$^c$ is switch controlling software which handles matching functions to route the packets to appropriate network devices. FR is a finite set of information to represents forwarding rules denoted as FR=\{$ r_{1},r_{2}, r_{3}. . .r_{n}$\}. `SwitchTrust' is an assignment function `SwitchTrust': W $\rightarrow$ Label, which assigns each switch with a label, more the number of labels in a switch is considered as a most trusted switch.

 \subsection{SDN Controller}
   SDN controller is a tuple C:= \{$M^{h}$, FC, $\delta$, $s_{0}^{c}$, $S^{c}$\} that manages control flow of the network based on OpenFlow protocol. 
    \begin{itemize}
    \item $S^{c}$ is a set of control states. 
    \item $s_{0}^{c}$ is a set of initial control states, $s_{0}^{c}$ $\subseteq$ $S^{c}$. 
    \item $M^{h}$: h$\rightarrow$$h^{'}$ is a header modification function that modifies the packet header based on controller policies. 
    \item FC: h $\times$ $N^{s}$ $\times$ PESA $\rightarrow$ FR  is a forward rule calculator function, where h is a packet header, $N^{s}$: $W^{s}$ $\times$  h$\rightarrow$ $\pi$  is a network status, which is used to find a feasible path of a packet in a set of all possible paths $\pi$. Paths $\pi$ are arranged based on feasibility to reach destination i.e. \{$\pi_{0}, \pi_{1}, \pi_{2}, . . . ,\pi_{n}$\}$\in$$\pi$, where  $\pi_{0}$ ranks high to route a packet, $\pi_{1}$ ranks next of $\pi_{0}$. \item $\delta$ $\subseteq$ $S^{c}$ $\times$  $M^{h}$ $\times$ FC $\times$ $S^{c}$ is a transition relation.
    \end{itemize}

\subsection{OpenFlow Table}
This section describes the components of the flow table `FTab' along with the mechanics of matching and action handling. OpenFlow switch has two types: OpenFlow only and OpenFlow- hybrid. OpenFlow switches support OpenFlow pipeline only. OpenFlow-hybrid switches support both OpenFlow operation and Ethernet switching operation.

The OpenFlow pipeline of every switch contains one or more flow tables, each flow table contains multiple entries. The OpenFlow pipeline processing unit defines how packets interact with the flow tables. An OpenFlow switch is required to have at least one ingress flow table (flow table to handle packet at an input port of a switch). Pipeline processing happens in two stages: ingress processing and egress processing (egress port that forwards network traffic to analysis tools). The outcome of ingress processing is to forward a packet to an output port, the switch may perform egress processing in the context of an output port. Egress processing is optional, and a switch may not support egress processing.  

During the processing of the flow table, a packet header is matched against the flow entries of the flow table. If a flow entry is found, the instruction in that flow entry is executed. If the listing is matched, the instructions in that flow entry are executed. These instructions are used to route the packet to another flow table if the current flow table priory is less than other. If the matching flow entry does not direct to another flow table, then the current flow entries are used to forward the packet to an output port. If a packet does not match with any flow entries in a flow table, then we call it table miss based on flow table configuration.  The instructions of a missed flow entry in a flow table can specify how to handle unmatched packets. The unmatched packet may drop, passing to another table or sending to a controller for further configuration.

The OpenFlow expects the mapping between network elements is consistent in following functionalities:
\begin{itemize}
\item Table consistency, in which the packet must match same as all other OpenFlow Tables. Furthermore, a difference in matching is due to contents of a flow table, and the packet header cannot be altered unless explicitly specified by the OpenFlow processing. 
\item Flow entry consistency, in which the actions of a flow entry apply to a packet is consistent with flow entry match. If a match field of flow entry matches a  packet header, then the set-field of a flow table modifies the packet header, unless explicit OpenFlow processing has modified the packet.      
\end{itemize}

A flow table entry of the form, \{match fields, priority, counters, instructions, timeouts, cookie, flags\}. A match field is to match packets, and it consists of the ingress port and packet header information, and optionally contains meta-data of the previous flow table. A priority for matching flow entry of a flow table. Counters are updated once the packet is matched with a flow entry.  An instruction is used to modify the action set and pipeline processing. Timeouts are the maximum amount of time or idle time before the switch expires flow. A cookie is opaque data chosen by a controller; this data is used by the controller to filter the flow entries affected by flow statistics, flow modification and deletion requests. A flag alters the flow entries, for example, the flag OFFF\_SEND\_FLOW\_REMOVAL is used to remove messages of that flow entry. All flow tables must support a table-miss flow entry to handle table misses. A table-miss entry has a piece of information to handle unmatched packets. An unmatched packet may be: dropped, send to a controller for further configuration or forwarded to subsequent flow tables.

\subsection{Topology and Configuration}
  Topology `T' of an SDN is a binary relation  T$\subseteq$((W $\times$ OP)$\times$(W $\times$ IP)) on switches and ports of the network system. The nodes of a network are represented by a relation (W $\times$ OP) and (W $\times$ IP). Furthermore, $<$h, p, w$>$ denotes the packet state in the network, where `h' is a packet header, and `p' and `w' is a port of a switch `w'.
     \begin{equation}
  Trans(nodes)=\begin{cases}
    \bigwedge\limits_{\exists w \in W} ConfiG_{w}(<h, p>, <h, p>), & \text{if $nodes<2$}.\\
   \bigwedge\limits_{w\in W}ConfiG_{w}(<h, p>, <h, p'>), & \text{if $nodes\geq2$}.
  \end{cases}
\end{equation}

  \begin{equation}
  Trans(nodes)=\begin{cases}
    \bigwedge\limits_{\exists w \in W} ConfiG_{w}(<h, p>, <h, p>), & \text{if $nodes<2$}.\\
   \bigwedge\limits_{w\in W}ConfiG_{w}(<h, p>, <h', p'>), & \text{if $nodes\geq2$}.
  \end{cases}
\end{equation}
 
Once the topology of the network is ready, then the packet transmission relation among the switches is given in recursive relation. Equation 1 presents a packet transmission relation without modifying packet header `h' from input port p to p'. Equation 2 presents a packet transmission relation with modifying packet header `h' from input port p to p'. In both equations 1 and 2, the first condition of the equation presents a packet dropping node, where the number of nodes in a network is not sufficient to send a packet across the network. The second condition of the equations presents a transmission relation across the nodes of a network based on network configuration function ConfiG.

 ConfiG is network configuration function, which assigns forwarding rules to switches, ConfiG: W $\rightarrow$ FR. Packets enter into an input port of a switch is configure and forwarded based on forwarding rules. A run of an SDN is a sequence of transmission nodes is present as:
 \begin{equation}
 run=(ConfiG_{0},w_{0}) \xrightarrow{P_{1}^{t}} (ConfiG_{1},w_{1}) \xrightarrow{P_{2}^{t}} . . .\newline \xrightarrow{P_{i}^{t}}(ConfiG_{i},w_{i})\xrightarrow{P_{i+1}^{t}}(*)
 \end{equation}
 
 Pair $(ConfiG_{i},w_{i})$ is considered as routing configuration by a controller to the switch $w_{i}$. A run is a sequence of nodes in a network, which allows packet $P_{i}^{t}$ is passing through in it.

\section{Representation of SDN in Kripke and SAT Structure for Model Checking}
 Verification of SDN security properties is essential, and it enhances the confidence of the system.  In this section, we present an SDN's Kripke structure along with specifications of SDN using LTL and CTL. Kripke structure of an SDN and its specifications are used in the verification process via model checking. Due to the advantages of Bounded model checking, we reduce the model checking problem with a bound, which can be solved by SAT solvers. 
 
 The Kripke structure of an SDN is a tuple SDN=(S$_{0}$, S, T, L) as shown in figure 1, where:

 \begin{itemize}
 \item S$_0$ is a set of initial states where, S$_0$$ \subseteq$ S.
    \item `S' is a set of states ($S^{c},S^{w}\in$S, where $S^{c}$ represents states belong to the controller and $S^{w}$ represents states belong to switch). 
 
 \item T $\subseteq$ S $\times$ S is a transition relation (\{($S^{c}$$\times$$S^{c}$), ($S^{c}$$\times$$S^{w}$), ($S^{w}$$\times$$S^{c}$), ($S^{w}$$\times$$S^{w}$) \} $\in$ T), where $S^{c}$ and $S^{w}$ is a set of states in a controller and switch.
 \item `L' is a labelling function `L': `S'$\rightarrow$ 2$^A$. In our model we used boolean vector $<$I, W, C , FC, M$^{h}$, O $>$ $\in$ A. `I' and `O' are the bits to represent the input and output port of a switch and a controller. `W' is a bit to represent a packet state in a switch and `C' is a bit to represent the packet state in a controller. FC and M$^{h}$ are the bits to represent states, where forward rules calculation and header modification happens. 
\end{itemize}

 \begin{figure}
  \centering
      \includegraphics[width=1.1\textwidth]{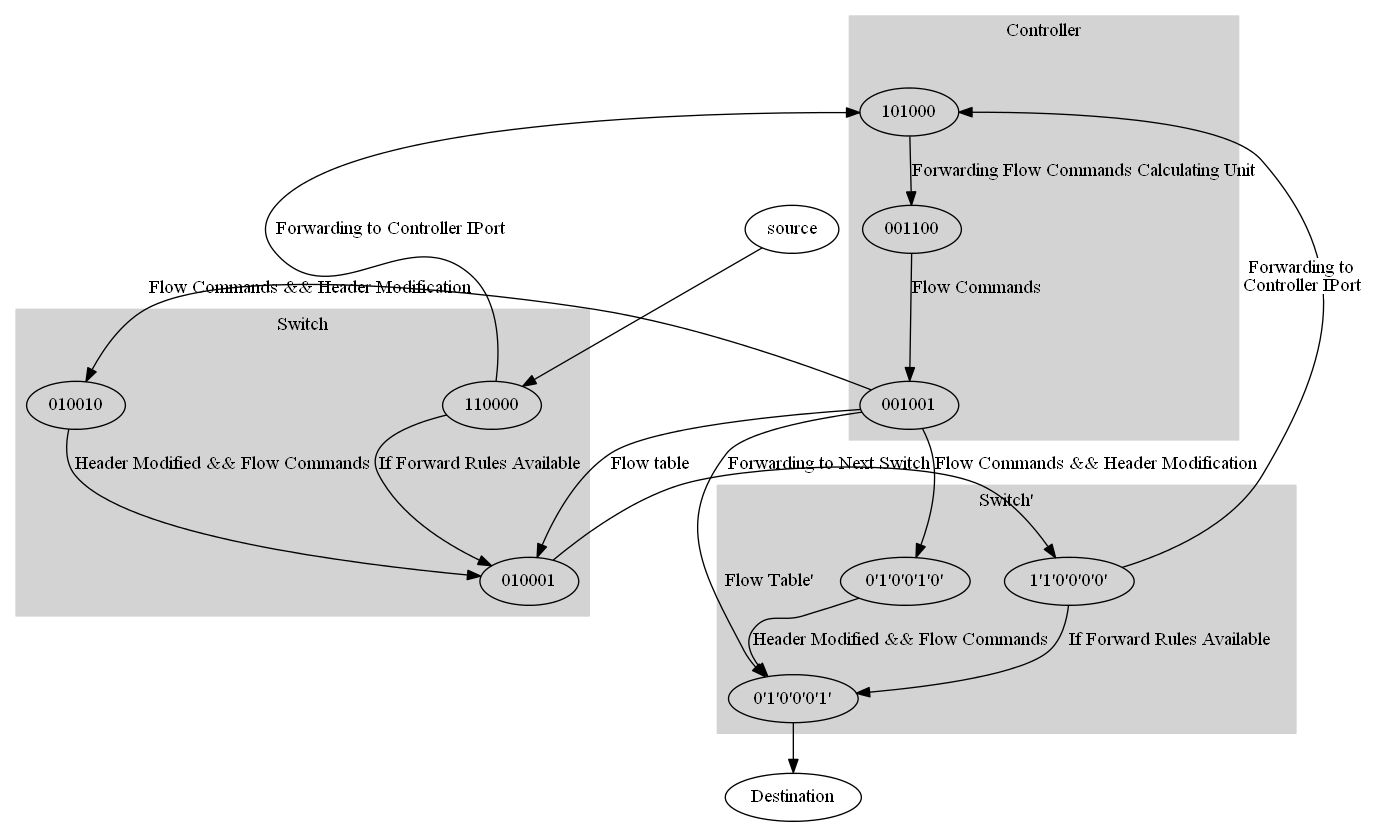}
  \caption{kripke model for software defined network}
\end{figure}
 
Given a Kripke structure of SDN, the specifications of a system are expressed using an LTL or CTL formula `f' and a bound `d', and semantics of BMC  can describe a  process of constructing a propositional formula [SDN, X$_{f}$]$_{d}$.

Let (s$_0$, s$_1$, s$_2$, . . s$_d$) be a finite sequence of states in a path $\pi$. The description of a formula  [SDN, X$_{f}$]$_{d}$ contains two components: SDN is a propositional formula that contains (s$_0$, s$_1$, s$_2$, . . s$_d$) and X$_{f}$ is also a propositional formula to validate the constraints given in a formula `f'. To define specification constraints X$_{f}$, we provide a definition of loop condition `L$^{j}$' which is propositional formula that is true if there is a loop in path $\pi$. Loop condition is considered as valid if there is a transition from `j' to previous states.

\begin{equation}
    L^{j} := \bigvee\limits_{i=0}^{j}(S_{j}, S_{i})
\end{equation}

For a Kripke structure SDN, and d $\geq$ 0,

\begin{equation}
    [SDN]_{d} := Init(S_{0}) \wedge \bigwedge\limits_{i=1}^{d-1}(S_{i}, S_{i+1})
\end{equation}
The complete formula, which includes an SDN model and specification is presented in a boolean formula [SDN, X$_{f}$]$_{d}$.

\begin{equation}
    [SDN, X_{f}]_{d} := Init(S_{0}) \wedge \hspace{2mm} \bigwedge\limits_{i=1}^{d-1}(S_{i}, S_{i+1}) \hspace{2mm} \wedge (\neg L^{d} \wedge [X_{f}]_{d})
\end{equation}

Consider a Kripke structure of an SDN in figure 2.  Each state of an SDN represented by five-bit variables. We use b[5], b[4], b[3], b[2], b[1] and b[0], where  b[5] is a high bit and b[0] be the low bit. The initial state of an SDN is represented as follows,
\begin{equation}
Init(S_{0}) := b[5] \wedge b[4] \wedge \neg b[3] \wedge \neg b[2] \wedge \neg b[1] \wedge \neg b[0]
\end{equation}
The transition relation of an SDN is represented as follows,

\begin{equation}
\begin{split}
T(s,s') := \Huge \{ (b[5] \wedge b[4] \wedge \neg b[3] \wedge \neg b[2] \wedge \neg b[1] \wedge \neg b[0] \\
\wedge b'[5] \wedge \neg b'[4] \wedge b'[3] \wedge \neg b'[2] \wedge \neg b'[1] \wedge \neg b'[0]) \\
\vee \\
(b[5] \wedge \neg b[4] \wedge b[3] \wedge \neg b[2] \wedge \neg b[1] \wedge \neg b[0] \\
\wedge \neg b'[5] \wedge \neg b'[4] \wedge b'[3] \wedge  b'[2] \wedge \neg b'[1] \wedge \neg b'[0])  \\
\vee \\
(\neg b[5] \wedge \neg b[4] \wedge b[3] \wedge b[2] \wedge \neg b[1] \wedge \neg b[0] \\
\wedge \neg b'[5] \wedge \neg b'[4] \wedge b'[3] \wedge \neg b'[2] \wedge \neg b'[1] \wedge b'[0]) \\
\vee \\
(\neg b[5] \wedge \neg b[4] \wedge b[3] \wedge \neg b[2] \wedge \neg b[1] \wedge b[0] \\
\wedge \neg b'[5] \wedge b'[4] \wedge \neg b'[3] \wedge \neg b'[2] \wedge \neg b'[1] \wedge b'[0])  \\
\vee \\
(\neg b[5] \wedge \neg b[4] \wedge b[3] \wedge \neg b[2] \wedge \neg b[1] \wedge b[0] \\
\wedge \neg b'[5] \wedge b'[4] \wedge \neg b'[3] \wedge \neg b'[2] \wedge b'[1] \wedge \neg b'[0])  \\
\vee \\
(\neg b[5] \wedge b[4] \wedge \neg b[3] \wedge \neg b[2] \wedge b[1] \wedge \neg b[0] \\
\wedge \neg b'[5] \wedge b'[4] \wedge \neg b'[3] \wedge \neg b'[2] \wedge \neg b'[1] \wedge b'[0])   \\
\vee \\
(\neg b[5] \wedge b[4] \wedge \neg b[3] \wedge \neg b[2] \wedge \neg b[1] \wedge b[0] \\ 
\wedge b'[5] \wedge b'[4] \wedge \neg b'[3] \wedge \neg b'[2] \wedge \neg b'[1] \wedge \neg b'[0])  \\
\vee \\
(b[5] \wedge b[4] \wedge \neg b[3] \wedge \neg b[2] \wedge \neg b[1] \wedge \neg b[0] \\
\wedge \neg b'[5] \wedge  b'[4] \wedge \neg b'[3] \wedge \neg  b'[2] \wedge \neg b'[1] \wedge b'[0]) \\
\}
\end{split}
\end{equation}

\subsection{Analyse with a Faulty Transition}
We now add a faulty transition from state 101000 to state 001001 denote by T$_{f}$.
\begin{equation}
\begin{split}
T_{f}(s, s') := T(s,s')
\vee
(b[5] \wedge \neg b[4] \wedge b[3] \wedge \neg b[2] \wedge \neg b[1] \wedge \neg b[0]
\wedge \neg b'[5] \wedge \neg b'[4] \wedge b'[3] \\ 
\wedge \neg b'[2] \wedge \neg b'[1] \wedge  b'[0])  
\end{split}
\end{equation}

Consider the primary property of an SDN, forward rules has to calculate for every packet which is pass through the controller for routing. The property is represented as  Gp, where p is $\neg b[5] \wedge \neg b[4] \wedge b[3] \wedge b[2] \wedge \neg b[1] \wedge \neg b[0] $. Using BMC, we generate a counterexample results witness of F$\neg$p. The absence of such property indicates the SDN property is violated.

 Consider a case where the bound d = 2 and unrolling the faulty SDN system transition relation in the following formula:
 \begin{equation}
[[SDN]]_{2} := Init(S_{0}) \wedge T_{f}(S_{0}, S_{1}) \wedge T_{f}(S_{1}, S_{2})
\end{equation}
The loop condition is represented as:
\begin{equation}
    L^{2} := \bigvee\limits_{i=0}^{2}(S_{2}, S_{i})
\end{equation}
The formula on a path without loops:
\begin{equation}
\begin{split}
[[F(\neg p)]]_{2}^{0} :=\neg p(S_{0}) \vee [[F(\neg p)]]_{2}^{1} \\
[[F(\neg p)]]_{2}^{1} :=\neg p(S_{1}) \vee [[F(\neg p)]]_{2}^{2} \\
[[F(\neg p)]]_{2}^{2} :=\neg p(S_{2}) \vee [[F(\neg p)]]_{2}^{3} \\
[[F(\neg p)]]_{2}^{3} := 0
\end{split}
\end{equation}
Finally by substituting all terms and we get:
\begin{equation}
[[F(\neg p)]]_{2}^{0} := \neg p(S_{0}) \vee \neg p(S_{1}) \vee \neg p(S_{2})
\end{equation}
By putting SDN transitions, loop condition and property together, we can get a new formula:
 \begin{equation}
[[SDN, F(\neg p)]]_{2} := [[SDN]]_{2} \wedge (\neg  L^{2} \wedge [[F(\neg p)]]_{2}^{0})
\end{equation}
Since a missing path is sufficient to prove a violation of SDN property, the loop condition is excluded. This results in the following formula: 
\begin{equation}
\begin{split}
[[SDN, F(\neg p)]]_{2} := [[SDN]]_{2} \wedge[[F(\neg p)]]_{2}^{0} := \\
Init(S_{0}) \wedge T_{f}(S_{0}, S_{1}) \wedge T_{f}(S_{1}, S_{2}) \wedge (\neg p(S_{0}) \vee \neg p(S_{1}) \vee \neg p(S_{2}))
\end{split}
\end{equation}
The assignment 110000, 101000, 001001 satisfies the above formula $[[SDN, F(\neg p)]]_{2}$; this assignment violates the fundamental property of an SDN.

\subsection{Expression of SDN Properties using Temporal Logic}
\begin{description}

\item [$AG(W_{ip} \rightarrow A(\neg W^{'}_{ip} \textbf{W} FC)$: ]For any state in SDN, it is not possible to send a packet to next switch without having forwarding rules. 
\item[$AG(C_{ip} \rightarrow AX(FC))$:] For any state in SDN, where a packet enters the controller always calculate forwarding rules for missed flow-entry before sending it to the switch. 
\item [$AG(W_{ip} \wedge \neg FC \rightarrow AX(C_{ip}))$:] For any state in SDN, where a packet enters the switch forwards to the controller for calculating/configuring  forwarding rules if the switch misses the flow entry. 
\item [$AG (W_{ip} \wedge FC \rightarrow AX(W_{op}))$:] For any state in SDN, where a packet enters the switch's input port should forward to an output port if there are forwarding rules available to route the packet. 
\end{description}

\section{Conclusion}
In cloud computing, software-defined network (SDN) gaining more attention due to its advantages in network configuration to improve network performance and network monitoring. It is essential to ensure the correctness of SDN due to secure data transmitting in it. In this paper, we present a work to apply formal techniques on the software-defined network for verification. The software-defined network is formally described using the Kripke structure and its properties are formally presented using temporal logic formulas. The SDN LTL or CTL properties are verified against the formal Kripke model to check the specifications meets its model. Furthermore, we also analysed with a faulty transition and verified faulty transition violates the fundamental properties of the SDN.

\bibliographystyle{unsrt}  


\end{document}